\def\be{\begin{equation}}
\def\ee{\end{equation}}
\def\ba{\begin{array}}
\def\ea{\end{array}}

\documentclass[aps,amsmath,amssymb,amsfonts,showpacs,pra]{revtex4}
\usepackage{graphicx}
\usepackage{epstopdf}

\def\qed{\leavevmode\unskip\penalty9999 \hbox{}\nobreak\hfill
     \quad\hbox{\leavevmode  \hbox to.77778em{%
               \hfil\vrule   \vbox to.675em%
               {\hrule width.6em\vfil\hrule}\vrule\hfil}}
     \par\vskip3pt}

\newtheorem{theorem}{Theorem}
\newtheorem{corollary}{Corollary}
\newtheorem{lemma}{Lemma}
\begin{document}

\title{A Family of Bipartite Separability Criteria Based on Bloch Representation of Density Matrices}

\author{Xue-Na Zhu$^{1}$}
\thanks{E-mail:jing\_feng1986@126.com.}

\author{Jing Wang$^{2}$}
\author{Gui Bao$^{1}$}
\author{Ming Li$^{2}$}
\author{Shu-Qian Shen$^{2}$}
\author{Shao-Ming Fei$^{3,4}$}

\affiliation{$^1$School of Mathematics and Statistics Science, Ludong University, Yantai 264025, China\\
$^2$College of the Science, China University of
Petroleum, Qingdao 266580, China\\
$^3$School of Mathematical Sciences, Capital Normal
University, Beijing 100048, China\\
$^4$Max-Planck-Institute for Mathematics in the Sciences, 04103 Leipzig, Germany}

\begin{abstract}

\noindent{\bf{Abstract}} We study the separability of bipartite quantum systems in arbitrary dimensions based on the Bloch representation of density matrices. We present two separability criteria for quantum states in terms of the matrices $T_{\alpha\beta}(\rho)$ and $W_{ab,\alpha\beta}(\rho)$ constructed from the correlation tensors in the Bloch representation. These separability criteria can be simplified  and detect more entanglement than the previous separability criteria. Detailed examples are given to illustrate the advantages of results.
\end{abstract}

\maketitle
\section{Introduction}
Quantum entanglement \cite{F,K,H,J,C} lies at the heart of quantum
information processing and quantum computation \cite{ma}.
The quantification of quantum entanglement has drawn much attention in the last decade. A prior question in the study of quantum entanglement is to determine whether a given quantum state is entangled or not. Denote $H_{M}$ and $ H_{N}$ the vector spaces with dimensions $M$ and $N$, respectively. A bipartite $M\otimes N$ state $\rho\in H_{M}\otimes H_{N}$ is said to be separable if it can be written as a convex sum of tensor products of the states of subsystems,
\begin{equation}\label{sep}
\rho=\sum_{i}p_i\rho^{i}_{M}\otimes \rho^{i}_{N},
\end{equation}
where $p_i\geq 0$ and $\sum_ip_i=1$. Otherwise $\rho$ is said to be entangled.

As a consequence, much efforts have been devoted to the so-called separability problem. The most well-known one is the positive partial transpose (PPT) criterion \cite{ph1,ph2}, which is both necessary and sufficient for low-dimensional systems $2\otimes2$ and $2\otimes3$. For high-dimensional states, the PPT criterion is only a necessary one. A variety of separability criteria have been proposed so far, such realignment criteria \cite{r1,r2}, covariance matrix criterion (CMC) \cite{151} and so on \cite{1,2,3,S}. In particular, much subsequent works \cite{VJ,shen,Liming,2020} have been devoted to finding necessary conditions for separability based on Bloch representation of density matrices.

In terms of the Bloch representation any quantum state  $\rho\in H_{M}\otimes H_{N}$ can be written as,
\begin{equation}\label{B}
\rho=\frac{1}{MN}\big(I_{M}\otimes I_{N}+\sum_{k=1}^{M^2-1}r_k\lambda^{M}_k\otimes I_{N}+\sum_{l=1}^{N^2-1}s_l I_{M}\otimes\lambda^{N}_l
+\sum_{k=1}^{M^2-1}\sum_{l=1}^{N^2-1}t_{kl}\lambda^{M}_k\otimes\lambda^{N}_l\big),
\end{equation}
where  $I_{i}$ $(i=M,N)$ denote the $i\times i$ identity matrix,
$\lambda^{M}_i$, $i=1,2,...,M^2-1$, are the generators of $SU(M)$ given by $\{\omega_l,u_{jk},v_{jk}\}$ with
$\omega_l=\sqrt{\frac{2}{(l+1)(l+2)}}\left(\sum_{i=0}^{l}|i\rangle\langle i|-(l+1)|l+1\rangle\langle l+1|\right)$, $u_{jk}=|j\rangle\langle k|+|k\rangle\langle j|$, $v_{jk}=-i(|j\rangle\langle k|-|k\rangle\langle j|)$, $0\leq l\leq M-2$ and $0\leq j< k\leq M-1$, $r_i=\frac{M}{2}Tr(\rho\lambda^{M}_i\otimes I_{N})$,
$s_i=\frac{N}{2}Tr(\rho I_{N}\otimes\lambda^{N}_i)$
and
$t_{ij}=\frac{MN}{4}Tr(\rho\lambda^{M}_i\otimes \lambda^{N}_j).$

Denote
$r=(r_1,...,r_{M^2-1})^t$ and $s=(s_1,...,s_{N^2-1})^t$, where $t$ stands for transpose.
Let $T(\rho)$ be the matrix with entries $t_{kl}$.
If the bipartite state $\rho\in H_M\otimes H_N$ with Bloch representation
(\ref{B}) is separable, it has been shown that \cite{VJ}
\begin{equation}\label{VJ}
||T(\rho)||_{KF}\leq \sqrt{\frac{MN(M-1)(N-1)}{4}},
\end{equation}
where the Ky Fan matrix norm is defined as the sum of the singular value of the matrix, $||A||_{KF}=Tr\sqrt{A^{\dagger}A}.$
In \cite{Liming} the authors presented a stronger separability criteria,
\begin{equation}\label{L}
||T^{'}(\rho)||_{KF}\leq \frac{\sqrt{(M^2-M+2)(N^2-N+2)}}{2MN}
\end{equation}
for separable states, where $T^{'}(\rho)=\begin{pmatrix}
1&s^t\\
r&T(\rho)
\end{pmatrix}.$
In \cite{shen}, the authors constructed the following matrix,
\begin{equation*}\label{shen}
S^m_{ab}(\rho)= \begin{pmatrix}
abE_{m\times m}&a w^t_{m}(s)\\
b w_{m}(r)&T(\rho)
\end{pmatrix},
\end{equation*}
where $a$ and $b$ are nonnegative real numbers, $E_{mm}$ is the $m\times m$ matrix with all entries being $1$, $m$ is a given natural number,
$w_m(x)$ denotes $m$ columns of the column vector $x$, i.e., $w_m(x)=(x...x).$
The Theorem 1 of \cite{shen} showed that if the state $\rho\in H_M\otimes H_N$ is separable, then $\rho$ satisfies
\begin{equation}\label{shenj}
||S^m_{ab}(\rho)||_{KF}\leq \frac{1}{2}\sqrt{(2ma^2+M^2-M)(2mb^2+N^2-N)},
\end{equation}
which is even stronger than the previous criteria.

\section{Separability Conditions from the Bloch Representation based on $T_{\alpha\beta}(\rho)$}

Denote $\alpha=(a_1,...,a_{n})^t$ and $\beta=(b_1,...,b_{m})^t$, where
$a_i$ $(i=1,...,n)$ and $b_j$ $(j=1,...,m)$ are given real numbers, $m$ and $n$ are positive integers. We define the following matrix,
\begin{equation}\label{Tab1}
T_{\alpha\beta}(\rho)=\begin{pmatrix}
\alpha\beta^t&\alpha s^t\\
r\beta^t&T(\rho)
\end{pmatrix}.
\end{equation}
Using $T_{\alpha\beta}(\rho)$, we have the following separability criterion for bipartite states.

\begin{theorem}\label{TH1}
If the state $\rho\in H_{M}\otimes H_{N}$ is separable, then
\begin{equation}\label{th1}
||T_{\alpha\beta}(\rho)||_{KF}\leq \sqrt{||\alpha||^2_2+\frac{M(M-1)}{2}}
\sqrt{||\beta||^2_2+\frac{N(N-1)}{2}},
\end{equation}
where $||\cdot||_2$ is the Euclidean norm on $R^{N^2-1}$.
\end{theorem}

{\sf [Proof]}~A bipartite quantum state with Bloch representation (\ref{B})
is separable if and only if there exist vectors $\mu_i\in R^{M^2-1}$ and
$\nu_i\in R^{N^2-1}$ with $||\mu_i||_2=\sqrt{\frac{M(M-1)}{2}}$ and
$||\nu_i||_2=\sqrt{\frac{N(N-1)}{2}}$, and $0<p_i\leq1$ with $\sum_ip_i=1$ such that
\begin{equation*}
T(\rho)=\sum_ip_i\mu_i\nu^t_i, r=\sum_ip_i\mu_i, s=\sum_ip_i\nu_i.
\end{equation*}
The matrix  $T_{\alpha\beta}(\rho)$ can then be written as,
\begin{eqnarray*}
T_{\alpha\beta}(\rho)&=&\begin{pmatrix}
\alpha\beta^t&\alpha s^t\\
r\beta^t&T(\rho)
\end{pmatrix}\\
&=&\sum_ip_i
\begin{pmatrix}
\alpha\beta^t&\alpha \nu^t_i\\
\mu_i\beta^t&\mu_i\nu^t_i
\end{pmatrix}\\
&=&\sum_ip_i\begin{pmatrix}
\alpha\\
\mu_i
\end{pmatrix}
\begin{pmatrix}
\beta^t, \nu^t_i
\end{pmatrix}.
\end{eqnarray*}
Therefore,
\begin{eqnarray*}
||T_{\alpha\beta}(\rho)||_{KF}
&\leq&\sum_ip_i\left|\left|\begin{pmatrix}
\alpha\\
\mu_i
\end{pmatrix}\right|\right|_2\cdot
\left|\left|\begin{pmatrix}
\beta^t, \nu^t_i
\end{pmatrix}\right|\right|_2\\
&=&\sqrt{||\alpha||^2_2+\frac{M(M-1)}{2}}
\sqrt{||\beta||^2_2+\frac{N(N-1)}{2}}.
\end{eqnarray*}
\hfill$\Box$

It can be seen that if we chose $a_i=a$ and $b_j=b$ for $i,j=1,...,n$
and $m=n$, Theorem \ref{TH1} reduces to the separability criterion (\ref{shenj}) given in \cite{shen}.

Define
\begin{equation*}\label{R}
R(\beta)=\begin{pmatrix}
p\beta\beta^{t}&\beta c^{t}\\
c \beta^{t}&\textrm{W}
\end{pmatrix},
\end{equation*}
where $p$ is a nonzero real number, $\beta$ ($c$) is a nonzero $n$ ($m$)-dimensional real vector, $\textrm{W}$ is an $m\times m$ Hermitian matrix.
We denote $\lambda_i(R(\beta))$ $(i=1,...,m+n)$ the singular values of $R(\beta)$ with $\lambda_i(R(\beta)) \leq \lambda_j(R(\beta))$ $(i\leq j)$.

\begin{lemma}\label{L1}
For $\beta_1\not=\beta_2$ but $||\beta_1||_2=||\beta_2||_2$, we have
$\lambda_i(R(\beta_1)) = \lambda_i(R(\beta_2))$ $(i=1,...,m+n).$
\end{lemma}

{\sf [Proof]}~
With respect to any nonzero real vector $\beta=(b_1,b_2,...,b_n)^t$, there exists a unitary matrix
$\textrm{U}$ such that $\textrm{U}\beta=(0,0,...,0,||\beta||_2)^t$. Then we have
\begin{equation*}
\begin{pmatrix}
U&0\\
0&I
\end{pmatrix}R\begin{pmatrix}
U^{\dagger}&0\\
0&I
\end{pmatrix}
=\begin{pmatrix}
0&0&0\\
0&p||\beta||^2_2&||\beta||_2 c^{t}\\
0&||\beta||_2c&\textrm{W}\\
\end{pmatrix}.
\end{equation*}
Denote \begin{equation*}\label{D}
D(\beta)=\begin{pmatrix}
p||\beta||^2_2&||\beta||_2 c^{t}\\
||\beta||_2c&\textrm{W}\\
\end{pmatrix}.
\end{equation*}
Since the singular values of an Hermitian matrix do not change under the unitary transformations, we have $\lambda_i(R(\beta)) = \lambda_i\left(\begin{pmatrix}
0&0\\
0&D(\beta)\\
\end{pmatrix}\right),$
$(i=1,...,m+n).$
Because of $D(\beta_1)=D(\beta_2)$, we complete the proof. \hfill$\Box$

Since the Ky Fan matrix norm
$||T_{\alpha\beta}(\rho)||_{KF}=Tr\sqrt{T_{\alpha\beta}(\rho)^{\dagger}T_{\alpha\beta}(\rho)}$,
$r\in \emph{R}^{M^2-1}$, $s\in \emph{R}^{N^2-1}$ and $T(\rho)\in\emph{R}^{(M^2-1)(N^2-1)}$, we have
\begin{equation*}
T_{\alpha\beta}(\rho)^{\dagger}T_{\alpha\beta}(\rho)=\begin{pmatrix}
(||\alpha||^2_2+||r||^2_2)\beta\beta^t&\beta(||\alpha||_2^2s^t+r^tT)\\
(||\alpha||_2^2s+T^tr)\beta^t&||\alpha||_2^2ss^t+T^t T
\end{pmatrix}.
\end{equation*}

By using Lemma 1 we have  the following corollary.

\begin{corollary}\label{jian}
For any quantum state $\rho$,
$||T_{\alpha\beta}(\rho)||_{KF}=||T_{||\alpha||_2||\beta||_2}(\rho)||_{KF}.$
\end{corollary}

From the Corollary \ref{jian}, we see that we only need to consider the norm of $\alpha$ and $\beta$ in dealing with the norm of $T_{\alpha\beta}(\rho)$. Hence, we simplify our Theorem \ref{TH1} to the following corollary.

\begin{corollary}\label{jian2}
If the state $\rho\in H_{M}\otimes H_{N}$ is separable, then
\begin{equation*}\label{jianhua}
||T_{ab}(\rho)||_{KF}\leq\sqrt{a^2+\frac{M(M-1)}{2}}
\sqrt{b^2+\frac{N(N-1)}{2}}
\end{equation*}
for any non-negative real numbers $a$ and $b$.
\end{corollary}

Corollary \ref{jian2} is equivalent to the Theorem \ref{TH1} with $||\alpha||_2=a$ and $||\beta||_2=b$.

{\it Example 1:} We consider the $2\otimes4$ state,
$
\rho_x=x|\xi\rangle\langle \xi|+(1-x)\rho,
$
where $|\xi\rangle=\frac{1}{\sqrt{2}}(|00\rangle+|11\rangle)$, $\rho$ is the bound entangled state considered in \cite{shen,Liming},
\begin{eqnarray*}
\rho&=\frac{1}{7d+1}&\begin{pmatrix}
d&0&0&0&0&d&0&0\\
0&d&0&0&0&0&d&0\\
0&0&d&0&0&0&0&d\\
0&0&0&d&0&0&0&0\\
0&0&0&0&\frac{1+d}{2}&0&0&\frac{\sqrt{1-d^2}}{2}\\
d&0&0&0&0&d&0&0\\
0&d&0&0&0&0&d&0\\
0&0&d&0&\frac{\sqrt{1-d^2}}{2}&0&0&\frac{1+d}{2}\\
\end{pmatrix}
\end{eqnarray*}
with $d\in(0,1)$. For simplicity, set $d=\frac{9}{10}$ and choose $\alpha=(\frac{1}{2\sqrt{3}},\frac{1}{2\sqrt{3}})^t$ and $\beta=(1,0)^t$.
Then Theorem \ref{TH1} detects the entanglement of $\rho_{x}$ for $x\in[0.223406,1]$.
One may also choose $\alpha=(a_1,..,a_n)^t$ and $\beta=(b_1,...,b_m)^t$ in general,
where $\sum_{i=1}^{n}a_i^2=\frac{1}{6}$ and $\sum_{i=1}^{m}b_i^2=1$. The result is the same.

Combining Theorem 1 and Corollary \ref{jian2}, we have the following theorem.

\begin{theorem}\label{C2}
If a state $\rho\in H_{M}\otimes H_{N}$ is separable, then
\begin{equation}\label{c2}
||T_{ab}(\rho)||_{KF}\leq \sqrt{\frac{NM(N-1)(M-1)}{4}}+
|ab|,
\end{equation}
where $a,b\in \textrm{R}$ and $|b|=|a|\sqrt{\frac{N(N-1)}{M(M-1)}}$.
\end{theorem}

{\sf [Proof]}~For a state $\rho\in H_{M}\otimes H_{N},$ we have
\begin{equation*}
||T_{ab}(\rho)||_{KF}=||\begin{pmatrix}
ab&as^t\\
 br&T(\rho)\\
\end{pmatrix}||_{KF}\geq |ab|+||T(\rho)||_{KF},
\end{equation*}
where the first inequality is due to
$
||\begin{pmatrix}
A&B\\
C&D\\
\end{pmatrix}||_{KF}\geq||A||_{KF}+||D||_{KF}
$
for any complex matrices $A, B, C$ and $D$ with adequate dimensions\cite{VJ}.
If $\rho$ is separable, we have
$$||T_{ab}(\rho)||_{KF}\leq \sqrt{a^2+\frac{M(M-1)}{2}}\sqrt{b^2+\frac{N(N-1)}{2}}
$$
and
$$
||T(\rho)||_{KF}\leq \sqrt{\frac{MN(M-1)(N-1)}{4}}.
$$
Setting $$\sqrt{a^2+\frac{M(M-1)}{2}}\sqrt{b^2+\frac{N(N-1)}{2}}=|ab|+\sqrt{\frac{MN(M-1)(N-1)}{4}},$$
we have $|b|=|a|\sqrt{\frac{N(N-1)}{M(M-1)}}.$
\hfill$\Box$

From the proof of Theorem 2, for the separable quantum states one has
 $$||T(\rho)||_{KF}\leq||T_{ab}(\rho)||_{KF}-|ab|
\leq \sqrt{\frac{MN(M-1)(N-1)}{4}}.$$
Theorem 2 can detect more entanglement than the Theorem 1 given in \cite{VJ},
see the following example.

{\it Example 2:} Consider the following bipartite qubit state, $\rho=p|\psi\rangle\langle\psi|+(1-p)|00\rangle\langle00|,$
where $p\in[0,1]$ and $|\psi\rangle=\frac{1}{\sqrt{2}}(|01\rangle+|10\rangle).$
Let $b=a\not=0$. We have $||T_{aa}(\rho)||_{KF}=2p+\sqrt{4a^2p^2+(2p-1-a^2)^2},$
which implies that $||T_{aa}(\rho)||_{KF}>1+a^2$ for $p\in(0,1]$.
Namely, the entanglement is detected for $p\in(0,1]$, which is better than the result $p\in(\frac{1}{2},1]$ from the Theorem 1 in \cite{VJ}.

\section{Separability Conditions from the Bloch Representation based on $W{ab,\alpha\beta}(\rho)$}

Next we define
\begin{equation}\label{Tab2}
W_{ab,\alpha\beta}(\rho)=\begin{pmatrix}
ab&a\alpha^t\otimes s^t\\
 b\beta\otimes r&\beta\alpha^t\otimes T(\rho)
\end{pmatrix},
\end{equation}
where $a$ and $b$ are real numbers.
Using $W_{ab,\alpha\beta}(\rho)$, we get the following separability criterion for bipartite states.

\begin{theorem}{\label{TH3}}
If the state $\rho\in H_{M}\otimes H_{N}$ is separable, then
\begin{equation}\label{th3}
||W_{ab,\alpha\beta}(\rho)||_{KF}\leq \sqrt{a^2+||\beta||^2_2\frac{M(M-1)}{2}}
\sqrt{b^2+||\alpha||^2_2\frac{N(N-1)}{2}},
\end{equation}
where $||\cdot||_2$ is the Euclidean norm on $R^{N^2-1}$.
\end{theorem}

{\sf [Proof]}~A bipartite quantum state with Bloch representation (\ref{B})
is separable if and only if there exist vectors $\mu_i\in R^{M^2-1}$ and
$\nu_i\in R^{N^2-1}$ with $||\mu_i||_2=\sqrt{\frac{M(M-1)}{2}}$ and
$||\nu_i||_2=\sqrt{\frac{N(N-1)}{2}}$, $0<p_i\leq1$ with $\sum_ip_i=1$ such that
$T(\rho)=\sum_ip_i\mu_i\nu^t_i$, $r=\sum_ip_i\mu_i$ and $s=\sum_ip_i\nu_i$.
Therefore, for separable states $\rho$ the matrix $W_{ab,\alpha\beta}(\rho)$ reduces to
\begin{eqnarray*}
W_{ab,\alpha\beta}(\rho)&=&\begin{pmatrix}
ab&a \alpha^t\otimes s^t\\
b\beta\otimes r&\beta\alpha^t\otimes T(\rho)
\end{pmatrix}\\
&=&\sum_ip_i
\begin{pmatrix}
ab&a \alpha^t\otimes \nu_i^t\\
b\beta\otimes\mu_i&\beta\alpha^t\otimes \mu_i\nu_i^t
\end{pmatrix}\\
&=&\sum_ip_i\begin{pmatrix}
a\\
\beta\otimes\mu_i
\end{pmatrix}
\begin{pmatrix}
b &\alpha^t\otimes\nu^t_i
\end{pmatrix}.
\end{eqnarray*}
Hence one gets
\begin{eqnarray*}
||W_{ab,\alpha\beta}(\rho)||_{KF}
&\leq&\sum_ip_i\left|\left|\begin{pmatrix}
a\\
\beta\otimes\mu_i
\end{pmatrix}\right|\right|_2\cdot
\left|\left|\begin{pmatrix}
b &\alpha^t\otimes\nu^t_i
\end{pmatrix}\right|\right|_2\\
&=&\sqrt{a^2+||\beta||^2_2\frac{M(M-1)}{2}}
\sqrt{b^2+||\alpha||^2_2\frac{N(N-1)}{2}},
\end{eqnarray*}
which proves the theorem. \hfill$\Box$

{\it Example 3:}
For the quantum state $\rho_x$ with $d=\frac{9}{10}$ in Example 1, if we take
$a=\frac{1}{\sqrt{6}}$, $b=1$, $\beta^t=(1,-2)$ and $\alpha^t=(1,3)$, Theorem \ref{TH3}
can detect the entanglement of $\rho_x$ for $x\in[0.22325,1]$,
which is better than the result $x\in[0.2234,1]$ from \cite{tinggui}.

Below we provide another example of PPT state whose entanglement is not detected by the filtered CMC \cite{151} but detected by our Theorem 3.

{\it Example 4:}
Consider a two qubit state,
\begin{eqnarray*}
\rho&=\frac{1}{2}&\begin{pmatrix}
1+a_1&0&0&a_3\\
0&0&0&0\\
0&0&a_2-a_1&0\\
t&0&0&1-a_2\\
\end{pmatrix},
\end{eqnarray*}
where the real parameters $\{a_1,a_2,a_3\}$ are taken such that $\rho\geq0.$
We choose $\alpha=(1,1)^t$, $\beta=(1,1)^t$, $a=\sqrt{2}x$ and $b=\sqrt{2}y$ in $W_{\alpha\beta}(\rho)$. From Theorem 3, we have that if $\rho$ is separable, then
\begin{equation}\label{t}
|a_3|+\sqrt{\lambda_{+}}+\sqrt{\lambda_{-}}\leq \sqrt{\frac{1+x^2}{2}}\sqrt{\frac{1+y^2}{2}},
\end{equation}
where\\
$\lambda_{\pm}
=\frac{1}{8}\left((1+a_1-a_2)^2+a_2^2x^2+a_1^2y^2+x^2y^2\pm
\sqrt{((1+a_1-a_2)^2+a_2^2x^2+a_1^2y^2+x^2y^2)^2-4(1+a_1)^2(1-a_2)^2x^2y^2}\right).$
The inequality (\ref{t}) is the same as the one from \cite{2020}, which recovers the $PPT$ condition for $\rho$.

Furthermore, we consider
the family of $3\otimes 3$ bound entangled states $\rho^x_{PH}$ introduced
by P. Horodecki\cite{P,S}.

{\it Example 5:} Consider the mixtures of $\rho^x_{PH}$ with the white noise,
$\rho(x,q)=q\rho^x_{PH}+(1-q)\frac{I}{9}$, where $0\leq q\leq1$ and
\begin{eqnarray*}
\rho^x_{PH}&=\frac{1}{8x+1}&\begin{pmatrix}
x&0&0&0&x&0&0&0&x\\
0&x&0&0&0&0&0&0&0\\
0&0&x&0&0&0&0&0&0\\
0&0&0&x&0&0&0&0&0\\
0&0&0&0&0&x&0&0&0\\
0&0&0&0&0&0&\frac{1+x}{2}&0&\frac{\sqrt{1-x^2}}{2}\\
0&0&0&0&0&0&0&x&0\\
x&0&0&0&x&0&\frac{\sqrt{1-x^2}}{2}&0&\frac{1+x}{2}\\
\end{pmatrix}.
\end{eqnarray*}
For simplicity, we let $x=0.9$. From the fig 4 of \cite{S},
$\rho(0.9,q)$  is entangled with $q>0.997$.
We take $a=\frac{1}{12}$, $b=\frac{1}{6}$, $\alpha=(\frac{1}{8},\frac{1}{8})^t$ and  $\beta=\frac{1}{8}$ in $W_{ab,\alpha\beta}(\rho(0.9,q))$. From our Theorem  3, $\rho(0.9,q)$  is entangled
when $q>0.9867$, which  is better than \cite{S}. See Fiq.1, where $\Delta=||W_{ab,\alpha\beta}(\rho(0.9,q))||_{KF}-\sqrt{a^2+3||\beta||^2_2}
\sqrt{b^2+3||\alpha||^2_2}.$

\begin{figure}[htpb]
\renewcommand{\figurename}{Fig.}
\centering
\includegraphics[width=6.5cm]{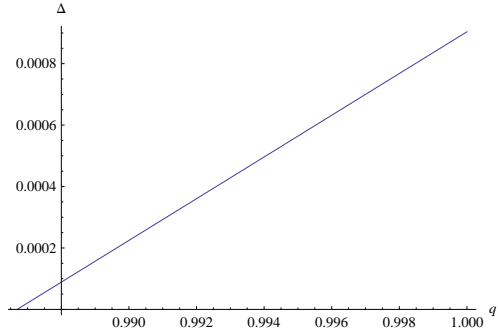}
\caption{{\small Entanglement detection of $\rho(0.9,q)$.}}
\label{Fig.1}
\end{figure}

Next, we give the relation Corollary \ref{jian2} and Theorem 3.

\begin{corollary}\label{w}
For quantum state $\rho\in H_A\otimes H_B$,
$||W_{ab,\alpha\beta}(\rho)||_{KF}=||W_{ab,||\alpha||_2||\beta||_2}(\rho)||_{KF}$
for any non-negative real numbers $a$ and $b$.
\end{corollary}

{\sf [Proof]}~
For $W_{ab,\alpha\beta}(\rho)$, we have
\begin{eqnarray}\label{www}
W^{\dag}_{ab,\alpha\beta}(\rho)W_{ab,\alpha\beta}(\rho)&
=&\begin{pmatrix}
ab^2+b||\beta||_2||r||_2&b\alpha^t\otimes(a^2 s^{t}+||\beta||_2r^tT)\\
b\alpha\otimes(a^2 s+||\beta||_2T^t r)&\alpha\alpha^t\otimes(a^2ss^t+||\beta||_2T^tT)\\
\end{pmatrix}.
\end{eqnarray}
From (\ref{www}), we have $||W_{ab,\alpha\beta}(\rho)||_{KF}=||W_{ab,\alpha||\beta||_2}(\rho)||_{KF}$.
For a given matrix $A$, one has $||A||_{KF}=Tr\sqrt{A^{\dag}A}=Tr\sqrt{AA^{\dag}}$.
Next, we have $||W_{ab,\alpha\beta}(\rho)||_{KF}=||W_{ab,||\alpha||_2\beta}(\rho)||_{KF}$ due to $W_{ab,\alpha\beta}(\rho)W^{\dag}_{ab,\alpha\beta}(\rho)$.
Then we obtain $||W_{ab,\alpha\beta}(\rho)||_{KF}=||W_{ab,||\alpha||_2||\beta||_2}(\rho)||_{KF}$.

For two positive numbers $k$ and $l$, we have
\begin{eqnarray*}\label{WWW}
W_{ab,kl}(\rho)&
=&\begin{pmatrix}
ab & ak s^{t}\\
bl r&kl T\\
\end{pmatrix}
=klT_{\frac{a}{l}\frac{b}{k}}(\rho).
\end{eqnarray*}

If the state $\rho\in H_{M}\otimes H_{N}$ is separable, from Corollary 2, we have
\begin{equation}\label{t11}
T_{\left(\frac{a}{l}\right)
\left(\frac{b}{k}\right)}(\rho)\leq\sqrt{\left(\frac{a}{l}\right)^2
+\frac{M(M-1)}{2}}\sqrt{\left(\frac{b}{k}\right)^2+\frac{N(N-1)}{2}},
\end{equation}
and from Theorem  3, we have
\begin{equation}\label{t12}
||W_{ab,kl}(\rho)||_{KF}\leq \sqrt{a^2+l^2\frac{M(M-1)}{2}}
\sqrt{b^2+k^2\frac{N(N-1)}{2}}.
\end{equation}
Form (\ref{t11}) and (\ref{t12}), one has Theorem 3 is equivalent to Corollary \ref{jian2} in detecting entanglement.

Note that the  family of bipartite separability criteria based on  $T_{\alpha\beta}(\rho)$ and $W_{ab,\alpha\beta}(\rho)$   come down to the Corollary \ref{jian2}, which is   only depend on real parameters $a$ and $b$.
Proposition 1 of Ref.\cite{shen} shown that the result of \cite{shen} becomes more effective when $m$ gets larger. From Corollary \ref{jian2} and the Proposition 1 of Ref.\cite{shen}, we know that Corollary \ref{jian2}  becomes more effective when $a$ and $b$ are selected large enough and  satisfy
$b\sqrt{M(M-1)}=a\sqrt{N(N-1)}$.

{\it Example 6:}
Let consider
 a  generalization of  well known $d_1\otimes d_2$ isotropic states\cite{2021}
\begin{equation}
\rho_{p}
=\frac{1-p}{d_1d_2}I_{d_1}\otimes I_{d_2}+p|\psi^{+}_{d_1}\rangle\langle\psi^{+}_{d_1}|,
\end{equation}
where $|\psi^{+}_{d_1}\rangle=\frac{1}{\sqrt{d_1}}\sum_{i=1}^{d_1}|e_i\otimes f_i\rangle$,
$|e_i\rangle$ defines orthonormal basis in $H_{d_1}$ and $|f_i\rangle$ defines orthonormal set in
$H_{d_2}$.

It is well known  that this state is separable if and only
if it is $PPT$ which is equivalent to $p\leq\frac{1}{d_2+1}$.
For simplicity, we take $d_1=2$ and $d_2=3$ for $\rho_{p}$ in the  example 6.
We show that Corollary \ref{jian2}  detects more entangled state than
de Vicente criterion\cite{VJ}, realignment criterion \cite{r1,r2} and  criterion based on
SIC POMVs(ESIC)\cite{S} for $\rho_{p}$.
And, we show that  Corollary \ref{jian2} becomes more effective when $a$ and $b$ get larger with $\frac{b}{a}=\sqrt{3}$.

We take $a=\sqrt{2}$ and $b=\sqrt{6}$ of  Corollary \ref{jian2} for $\rho_{p}$.
Then Corollary \ref{jian2} can detect the entanglement in $\rho_{p}$ for $p\geq0.378054$,
while the de Vicente criterion,
realignment criterion and ESIC criterion
 can only detect the entanglement in $\rho_{p}$ for $p\geq0.3849$, $p\geq0.3846$
 and $p\geq0.3819$, respectively.
At last, we choose $a=\sqrt{2}t$ and $b=\sqrt{6}t$ with $t>0$,
Then Corollary \ref{jian2} can detect the entanglement in
 $\rho_{p}$ for
 $p\geq0.379712$ with $t=\frac{1}{10}$,
$p\geq0.378139$ with $t=\frac{1}{2}$,
$p\geq0.378032$ with $t=2$,
$p\geq0.378025$ with $t=10$, respectively.

\section{Conclusions and Remarks}

In summary, based on the Bloch representation of a bipartite quantum state $\rho$, we have introduced the matrices $T_{\alpha\beta}(\rho)$ and show that
$||T_{\alpha\beta}(\rho)||_{KF}=||T_{||\alpha||_2||\beta||_2}(\rho)||_{KF}.$
i.e., the value of $||T_{\alpha\beta}(\rho)||_{KF}$  only depends on the norm of $\alpha$ and $ \beta$.  Thus the Theorem \ref{TH1} is equivalent to the Theorem 1 of \cite{shen} and can be further simplified to the Corollary \ref{jian2} which has a very simpler form. Meanwhile we have shown that the Corollary \ref{jian2} is more effective than the existing formula (\ref{VJ}). In addition, we have presented a separability criteria based on $W_{ab,\alpha\beta}(\rho)$, and show that $||W_{ab,\alpha\beta}(\rho)||_{KF}=||W_{ab,||\alpha||_2||\beta||_2}(\rho)||_{KF}$.  i.e., the value of $||W_{ab,\alpha\beta}(\rho)||_{KF}$  only depends on the norm of $\alpha$ and $ \beta$ for given $a$ and $b$.
At last, the  three separability criteria:  Theorem 1  of \cite{shen}, Theorem 1 and Theorem 3
can be  simplified to the Corollary \ref{jian2} which has a very simpler form.

\bigskip
\noindent{\bf Acknowledgments and Data  Availability Statements}\, \,
This work is supported by the Research Award Fund for Natural Science Foundation of Shandong Province No. ZR2021LLZ002, National Natural Science Foundation
of China under grant Nos. 12075159 and 12171044, Beijing Natural Science Foundation (Z190005), Academy for Multidisciplinary Studies, Capital Normal University, Shenzhen Institute for Quantum Science and Engineering, Southern University of Science and Technology (SIQSE202001), and the Academician Innovation Platform of Hainan Province. All data generated or analysed during this study are included in this published article.

\noindent{\bf Conflict of interest statement}\, \,
The authors declared that they have no conflicts of interest to this work.

\noindent{\bf{References}}

\end{document}